\documentclass[12pt]{iopart}
\usepackage{graphicx}
\usepackage{epstopdf}
\usepackage{iopams}
\usepackage{citesort}

\begin{document}

\title{Shape of the Growing Front of Biofilms}
%\title{Instability in the Shape of the Growing Front of Biofilms}

\author{Xin Wang$^1$, Howard A. Stone$^2$ and Ramin Golestanian$^1$ }

\address{$^1$ Rudolf Peierls Centre for Theoretical Physics, University of Oxford, Oxford, OX1 3NP, United Kingdom

$^2$ Department of Mechanical and Aerospace Engineering, Princeton University, Princeton, NJ 08544, USA}
\ead{ramin.golestanian@physics.ox.ac.uk}
\vspace{10pt}

\begin{abstract}
The spatial organization of bacteria in dense biofilms is key to their collective behaviour, and understanding it will be important for medical and technological applications. Here we study the morphology of a compact biofilm that undergoes unidirectional growth, and determine the condition for the stability of the growing interface as a function of the nutrient concentration and mechanical tension. Our study suggests that transient behaviour may play an important role in shaping the structure of a biofilm.
\end{abstract}

\section{Introduction}

The stability of a uniform front to small disturbances is a framework for understanding
pattern formation in many physical and biological systems \cite{RN335}, with a well-known example in material science being the fingering pattern formed due to supercooling of an alloy, as first characterized by Mullins and Sekerka \cite{RN367}. %pitt \cite{RN340}
In contrast, the self-organization and collective behaviour of living and synthetic active systems have been intensely studied in recent years \cite{RN325,RN350,RN321,RN342,RN349,RN372}, with an area of specific focus being spatial patterns generated by microbial systems \cite{RN337,RN324,RN333,RN329,RN343,RN344,RN341,RN327,RN328}. One particular example in the biological sciences that is receiving much recent attention concerns the growth and spatial structure of biofilms, which are densely packed bacterial communities \cite{RN357,RN358,RN359,RN360,RN361}.

Bacteria have been experimentally observed to form different patterns in the form of growing colonies when cultured on agar plates at different levels of nutrient concentration \cite{RN333,RN346,RN347,RN355,RN356}. Specifically, the surface  of growing colonies form circular (or flat) patterns when the nutrient concentration is high, while the patterns are fractal (or rough) when the nutrient concentration is low. The pattern formation driven by nutrient availability has been theoretically studied \cite{RN332,RN348,RN351,RN352,RN353} using various models such as the Fisher-Kolmogorov equation \cite{RN330,RN354}, which combines bacterial diffusion, bacterial growth and nutrient diffusion, all in the dilute limit. Recent studies have also highlighted the importance of the mechanical interactions between the cells  \cite{RN334,RN331}.
%rather than the bacterial diffusion

Given the wide diversity of microbial systems and their impact on both medical and natural systems, it is important to provide quantitative guidelines for instabilities that may influence the three-dimensional structure of growing biofilms. In this paper we present steps in this direction by analyzing the influence of two measurable quantities, namely the nutrient concentration and the effective surface tension that results from active mechanical interactions between bacteria \cite{RN371}, on the instability of a planar growing front of bacteria.

The shape instabilities introduced by nutrient factors have been recently studied using numerical simulations \cite{RN333,RN351,RN334,RN326,RN339}, which capture the main features of patterning in the experiments.
In this paper, we study the stability of the growing front of biofilms in a unidirectional planar growth using a perturbative analysis. We delineate various growth and patterning behaviours as a function of two key control parameters. With the mathematical criteria for the stability analysis, our study illustrates when and how the growing front of biofilms becomes unstable. Our analysis agrees with experimental studies concerning patterning of microbial colonies and can illustrate puzzles that are not fully understood in the previous simulation studies. Our study will be relevant to a wide variety of practical questions such as the behaviour of multispecies biofilms, the impact of digestive enzymes that may free nutrients, and the influence of cooperation among cells or the presence of cheater cells on the evolution of a biofilm. Moreover, it is inherently related to the recent stability analysis that has been used to study the chemically driven growth and division of droplets \cite{RN365} and may shed light on our understanding of division of proto-cells in early forms of life \cite{RN365,RN368}.

\section{Description of the System}

Consider the growth of a biofilm made of a single bacterial species. The scenario for culturing the system is depicted in Figure 1(a), where nutrient is supplied from the top of the domain. Denote the nutrient concentration as $c(x,y,z,t)$, where $x$, $y$, $z$ are the spatial coordinates and $t$ the time coordinate. The density of bacteria is $\rho  = 1/{b^3} $, with $b$ the characteristic length of a single bacterium. Within the biofilm, nutrient is consumed at a rate ${k}( c )$ by each cell, where $k(c) = k'\frac{c}{{c + {K_m}}}$ is a Michaelis-Menten form. This is a nonlinear form that describes crossover from a reaction-limited regime where the nutrient is abundant to a diffusion-limited regime where nutrient is scarce. While at the top layer of the growing biofilm, any of these regimes could be dominant, depletion of nutrient by every layer necessitates that at some depth there will be a crossover to the diffusion-limited regime. For the convenience of analysis, we apply $k( c ) \approx {k_0}c$ everywhere as an approximation. Because a diffusion process is involved, the nutrient concentration satisfies the equation:
\begin{equation}
{\partial _t}c - D{\nabla ^2}c + \rho   {k}( c ) \,  \theta ( { - z + {L_H}( {\mathbf{x},t} )} ) = 0,
\end{equation}
where $D$ denotes the diffusion coefficient, $\theta ( x )$ is the Heaviside step function, and ${L_H}( {\mathbf{x},t} )$ is the biofilm surface at position $\mathbf{x}=(x,y)$ and time $t$ (Figure 1(a)).

Denote the velocity of the growing front as  ${V}( {\mathbf{x},t} )$, then
\begin{equation}
{L_H}( {\mathbf{x},t} ) = \int_0^t {dt'V( {\mathbf{x},t'} ) + } {L_H}( {\mathbf{x},0} ).
\end{equation}
Supposing that ${N_u}$ nutrient molecules are consumed on average to make a single bacterium, we can calculate the velocity as
\begin{equation}
V( {\mathbf{x},t} ) = \frac{1}{{{N_{u}}}}  \int\limits_{( {{L_H}( {\mathbf{x},t} ) - H} )\,  \theta ( {{L_H}( {\mathbf{x},t} ) - H} )}^{{L_H}( {\mathbf{x},t} )} {dz\,{k}( c )},
\end{equation}
where $H$ is the depth of the active growing region within the biofilm (Figure 1(a)). The nutrient concentration and flux should be continuous at the biofilm surface. Nutrient diffusion is subject to boundary conditions
\begin{equation}
\left\{ \begin{array}{l}
c\left| {_{z = L}} \right. = {C_\infty },\\
c\left| {_{z = {L_H}^ + }} \right. = c\left| {_{z = {L_H}^ - }}, \right.\\
{\partial _z}c\left| {_{z = {L_H}^ + }} \right. = {\partial _z}c\left| {_{z = {L_H}^ - }}, \right.
\end{array} \right.
\end{equation}
where ${L_H}^ + $ denotes the boundary just above the biofilm surface and ${L_H}^ - $ the boundary below the surface. In the initial state, the nutrient concentration is homogenous within the culturing system, i.e. $c\left| {_{t = 0}} \right. = {C_\infty }$ (Figure 1(a)).

\begin{figure}[htb!]
\centering
\includegraphics[width=0.65\linewidth]{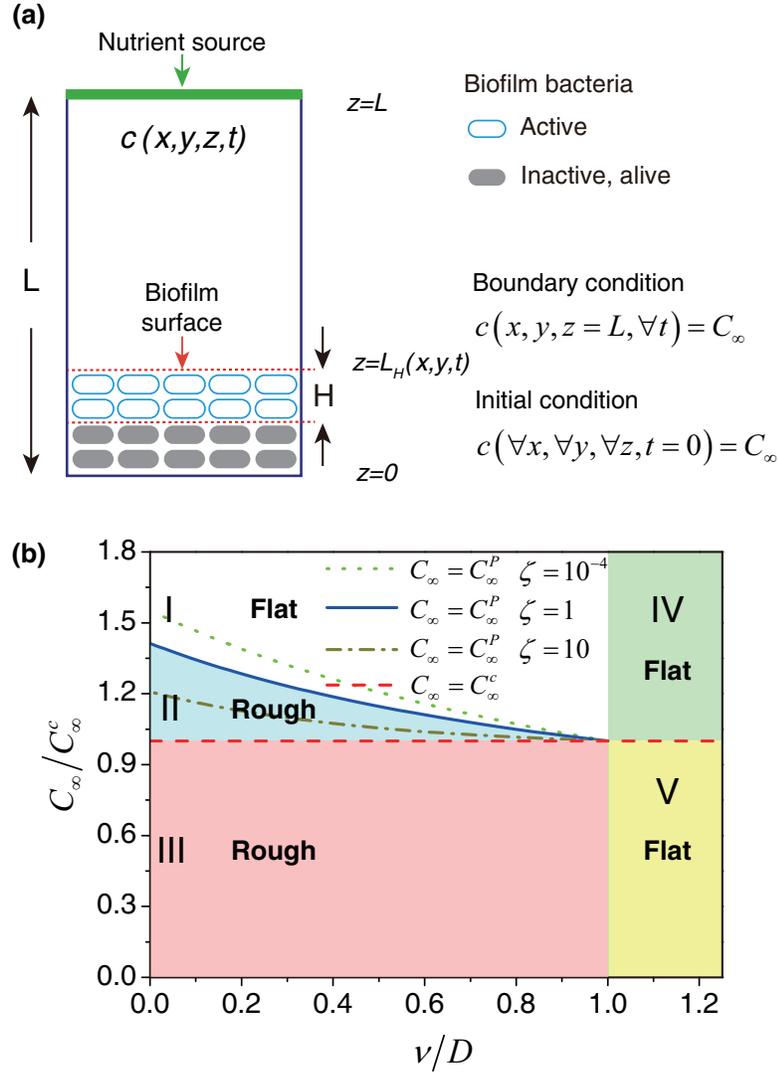}
\caption{
Surface shape of a biofilm determined by nutrient conditions and the effective surface tension coefficient.
(a) A schematic of bacterial growth of a biofilm. We assume exponential growth within a depth of $H$ below the biofilm surface, while those cells below this region belong to stationary phase (nutrient is still consumed, but there is no net cell growth). Nutrient is constantly supplied from the top of the domain to maintain $c(z=L)=C_\infty$. In the initial state, nutrient is distributed uniformly at concentration $C_\infty$. (b) Phase diagram regarding the shape of the biofilm surface ($\nu$ is the surface coefficient, $\zeta  \equiv {H^2}\rho {k_0}/D$; $\zeta=1$ is utilized for region partition). Regions I (white), IV (light green)and V (light yellow) are flat surface, while regions II (light blue)and III (pink) end up with a rough surface.}
\end{figure}

To faciliate analysis, we approximate the system as semi-infinite in the $z$-direction, and then we transform to a moving reference frame, $(x',y',z',t')$, with $(x',y',z',t')=(x,y,z - {L_H},t)$ and $c'(x',y',z',t')=c(x,y,z,t)$. In this case, the nutrient concentration satisfies the equation
\begin{equation}
{\partial _{t'}}c' - V{\partial _{z'}}c' - D{\nabla ^2}c' + \rho {k}(c') \,\theta ( { - z'} ) = 0.
\end{equation}

\section{Growth of a Flat Front}
Let us assume that the front grows in a steady state.  For a uniform stationary moving front, the one-dimensional description is (here $k( c ) \approx {k_0}c$)
\begin{equation}
 - V  \frac{{dc'}}{{dz'}} - D\frac{{{d^2}c'}}{{dz{'^2}}} + \rho   {k_0}c' \, \theta ( { - z'} ) = 0.
\end{equation}
Combined with the boundary conditions, the solution can be determined (see appendix A for details):
\begin{equation}
\begin{array}{lcr}
c'(z') = \left\{ \begin{array}{ll}
{C_\infty } \left(1 + B \,{e^{{{ - Vz'} \mathord{\left/
 {\vphantom {{ - Vz'} D}} \right.
 \kern-\nulldelimiterspace} D}}}  \right)& \; (z' \ge 0)\\
{C_\infty } \left(1 + B\right)\,{e^{{\lambda _1}z'}}& \;  (z' < 0)
\end{array} \right.\;;&{C_\infty } \ge  C_\infty ^c,\\
\texttt{\textrm{and}}\\
c'(z') = \left\{ \begin{array}{ll}
{C_\infty }\,\frac{{z'}}{L} &\;\;\;\;\;\;\;\;\;\;\;\;\;\;\;\;\;\;\; (z' \ge 0)\\
0 &\;\;\;\;\;\;\;\;\;\;\;\;\;\;\;\;\;\;\;(z' < 0) \\
\end{array} \right.\;;&{C_\infty } \le C_\infty ^c.\\
\end{array}
\end{equation}
where the coefficients are
\numparts
\begin{eqnarray}
C_\infty ^c = \frac{{{N_u}\rho }}{{1 - \exp ( - \sqrt \zeta  )}}\\
B =  - \frac{{1}}{\zeta }\,{\ln ^2}(1 - {N_u}\rho /{C_\infty })\\
%{C_1} = {C_\infty }\left[1 - \frac{1}{\zeta }{\ln ^2}(1 - {N_u}\rho /{C_\infty })\right]\\
{\lambda _1} =  - \frac{1}{H}\,\ln (1 - {N_u}\rho /{C_\infty })\\
\zeta  = {H^2}\rho {k_0}/D
\end{eqnarray}
\endnumparts
Meanwhile, the front velocity has a solution of the form:
\begin{equation}\displaystyle
V=\left\{
	\begin{array}{lr}
		\frac{D}{H}\left[ {\ln \left( {1 - \frac{{{N_u}\rho }}{{{C_\infty }}}} \right) - \frac{\varsigma }{{\ln \left( {1 - \frac{{{N_u}\rho }}{{{C_\infty }}}} \right)}}} \right]  & ;  \; {C_\infty } \ge  C_\infty ^c, \\\\
		0 &  ;  \;{C_\infty } \le  C_\infty ^c.\\
	\end{array}
\right.
\end{equation}

The velocity in steady state is zero when the nutrient concentration $C_\infty<C_\infty ^c$, while $V$ and $C_\infty$  is bijective when above the threshold $C_\infty ^c$. Equation (9) is plotted in Figure 2 for three different values of $\zeta$.
\begin{figure}[htb!]
\centering
\includegraphics[width=0.5 \linewidth]{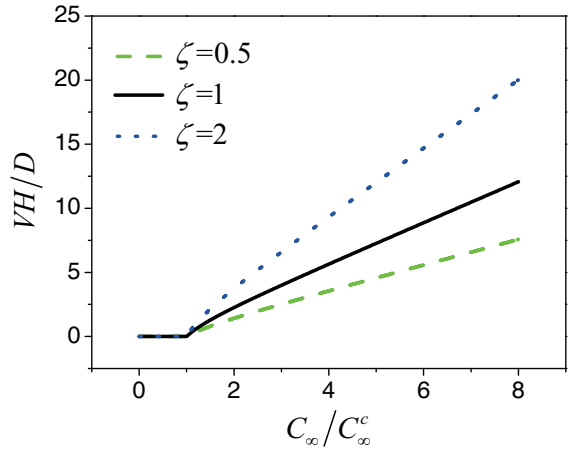}
\caption{Biofilm front velocity in steady state as a function of nutrient concentration.}
\end{figure}

\section{Growth Rate for Deformation Modes}
When $C_\infty >  C_\infty ^c$, and thus $V(t\to \infty)>0$ , consider the case that a growing front is slightly perturbed from the flat geometry, with the deformation described by a height profile function $h(\mathbf{x},t)$. For convenience, we define ${C^{\rm{\pm}}}$ as $c'$ above (+) or below (-) the biofilm surface. Then, the boundary conditions at the biofilm surface are
\begin{equation}
\left\{ \begin{array}{l}
{C^ + }\left| {_{z' = h( {\mathbf{x},t} )}} \right. = {C^ - }\left| {_{z' = h( {\mathbf{x},t} )}}, \right.\\
{\partial _{z'}}{C^ + }\left| {_{z' = h( {\mathbf{x},t} )}} \right. = {\partial _{z'}}{C^ - }\left| {_{z' = h( {\mathbf{x},t} )}}. \right.
\end{array} \right.
\end{equation}
We can construct a general solution for equation (5) of the form
\begin{equation}
\left\{ \begin{array}{l}
{C^{\rm{ + }}} = C_0^{\rm{ + }} + \int_{\mathbf q } {{A_+ }( {\mathbf q,t} )\;{{\mathop{\rm e}\nolimits} ^{i\mathbf q  \cdot  \mathbf x }}  } \;{{\mathop{\rm e}\nolimits} ^{ - {\alpha _ + }( \mathbf q )  z}},  \\
{C^ - } = C_0^ -  + \int_{\mathbf q } {{A_- }( {\mathbf q,t} )\;{{\mathop{\rm e}\nolimits} ^{i\mathbf q  \cdot  \mathbf x }}  }\; {{\mathop{\rm e}\nolimits} ^{{\alpha_ - }(\mathbf q )  z}},
\end{array} \right.
\end{equation}
where $C_0^ +  \equiv {C_\infty } \left(1 + B \, {e^{ - Vz'/D}} \right)$, $C_0^ -  \equiv {C_\infty } \left(1 + B \right)\,{e^{{\lambda _1}z'}}$, and $\int_{\mathbf q } { \equiv \int {\frac{{{d^2}{\mathbf q }}}{{{{( {2\pi } )}^2}}}} } $. By substituting equation (11) into equation (5), we find ${A_ \pm }( {\mathbf q,t} )$ and ${\alpha _ \pm }( \mathbf q )$ satisfy the following equations:
\begin{equation}
\left\{ \begin{array}{l}
{\partial _{t}}{A_ + } + ( {V{\alpha _ + } - D{\alpha _ + }^2 + D{q^2}} )\;{A_ + } = 0,\\
{\partial _{t}}{A_ - } - ( {V{\alpha _ - } + D{\alpha _ - }^2 - D{q^2} - \rho {k_0}} )\;{A_ - } = 0.
\end{array} \right.
\end{equation}
Combined with equation (10), one finds that ${A_ \pm }( {\mathbf q,t} )$ is of order $h$. Approximating these equations to the first order of $h$, we get the Fourier coefficients as
\begin{equation}
{A_ + } = {A_ - } =  - \frac{{{{{C_\infty } \left(1 + B \right)\,\rho {k_0}} \mathord{\left/
 {\vphantom {{{C_1}\rho {k_0}} D}} \right.
 \kern-\nulldelimiterspace} D}}}{{{\alpha _ + } + {\alpha _ - }}}\;h( {\mathbf q,t} ).
\end{equation}
As deformation is involved, there are two sources of contributions to the local front velocity $v( {\mathbf{x},t} )$,
\begin{equation}
v( {\mathbf{x},t} )=v_J( {\mathbf{x},t} )+v_b( {\mathbf{x},t} ),
\end{equation}
where $v_J( {\mathbf{x},t} )$ represents the biofilm growth caused by nutrient flux,
\begin{equation}
v_J( {\mathbf{x},t} )= \frac{1}{{{N_u}\rho }}\int\limits_{h( {\mathbf x,t} ) - H}^{h( {\mathbf x,t} )} dz'{( {V{\partial _{z'}}{C^ - } + D\partial _{z'}^2{C^ - }} )},
\end{equation}
while $v_b( {\mathbf{x},t})$ is a surface related contribution (a new source) that resists deformation in the growing front, as might be expected owing to cell-cell adhesion or the influence of type IV pili \cite{RN371}. For simplicity, we use the following generic form
\begin{equation}
{v_b}( {\mathbf x ,t} ){\rm{ = }}\nu   {\nabla ^2}h( {\mathbf x ,t} ),
\end{equation}
where $\nu$ is the effective surface tension coefficient. Developing an approximation to the first order of $h$ in equation (14) yields
\begin{equation}
\begin{array}{l}%\displaystyle
{v_J}\simeq {V}  \left[ {1 + \lambda _1  h( {\mathbf x ,t} )} \right] +
 \frac{V}{{{N_u}\rho }}  \int_{\mathbf q } {{A_ - }\;{{\mathop{\rm e}\nolimits} ^{i\mathbf q \cdot  \mathbf x }}  } ( {1 - {{\mathop{\rm e}\nolimits} ^{ - {\alpha _ - }  H}}} )  \\
 \;\;\;\;\;\;\;\;\;+ \frac{D}{{{N_u}\rho }} \int_{\mathbf q } {{A_ - }\;{{\mathop{\rm e}\nolimits} ^{i\mathbf q  \cdot \mathbf x }}  } ( {1 - {{\mathop{\rm e}\nolimits} ^{ - {\alpha _ - }  H}}} )  \;{\alpha _ - }.
\end{array}
\end{equation}
Meanwhile ${L_H}( {\mathbf x,t} )=\overline {{L_H}( t )} +h( {\mathbf x,t} )$, where $\overline {{L_H}( t )}$ denotes the average over $x$ and $y$ axis in ${L_H}( {\mathbf x,t} )$. By applying ${\partial _t}$ on both sides:
\begin{equation}
v( {\mathbf x ,t} ) = V + {\partial _t}h( {\mathbf x ,t} ).
\end{equation}
Combined with equations (14)-(17), in the Fourier space, we find that
\begin{equation}
{\partial _t}h( {\mathbf q ,t} ) = \left[ {\lambda (\mathbf q ) - \nu   {q^2}} \right]  h( {\mathbf q ,t} ),
\end{equation}
where
\begin{equation}
\lambda (\mathbf q ) = V  \lambda _1 - \frac{{V{\lambda _1}\left( {V + D{\alpha _ - }} \right)}}{{D( {{\alpha _ + } + {\alpha _ - }} )}}\frac{{( {1 - {{\mathop{\rm e}\nolimits} ^{ - {\alpha _ - }H}}} )}}{{( {1 - {{\mathop{\rm e}\nolimits} ^{ - {\lambda _1}H}}} )}}.
\end{equation}
Combining equation (19) with equations (12) and (13), we find that the unidentified functions ${\alpha _ \pm }(\mathbf  q )$ are subject to the following restrictions:
\numparts
\begin{eqnarray}
\lambda  - \nu   {q^2} =  D  {\alpha _ + }^2- V  {\alpha _+ }- D  {q^2}, \\
\lambda - \nu   {q^2} =  D  {\alpha _ - }^2+V  {\alpha _ - }- D  {q^2} - \rho   {k_0}.
\end{eqnarray}
\endnumparts
Equations (20) and (21) are in a closed form, from which we can obtain ${\alpha _ \pm }(\mathbf  q )$. Furthermore, from equation (19) it is clear that the stability of mode $q$ in the growing front is determined by the sign of ${\lambda (\mathbf q ) - \nu   {q^2}}$: when ${\lambda (\mathbf q ) - \nu   {q^2}}>0$, the deformation mode increases with time, and thus leads to an instability. Consequently, a growing front is stable only under the condition that there is no unstable mode, i.e. ${\lambda (\mathbf q ) - \nu   {q^2}}\leq0$ for all $q$. The dependence of instability on $q$ and $C_\infty$ is shown in Figure 3 with different values of $\zeta $ and ${\nu  \mathord{\left/
 {\vphantom {\nu  D}} \right.
 \kern-\nulldelimiterspace} D}$. One can find that in stable regions (Figure 3), $\lambda  - \nu   {q^2}$ peaks at ${{q} \approx 0} $. Thus, we can obtain the stability behavior (and thus the shape) of the growing front via the analysis of small $q$ regimes $(q \approx 0)$.
\begin{figure}[htb!]
\centering
\includegraphics[width=0.9\linewidth]{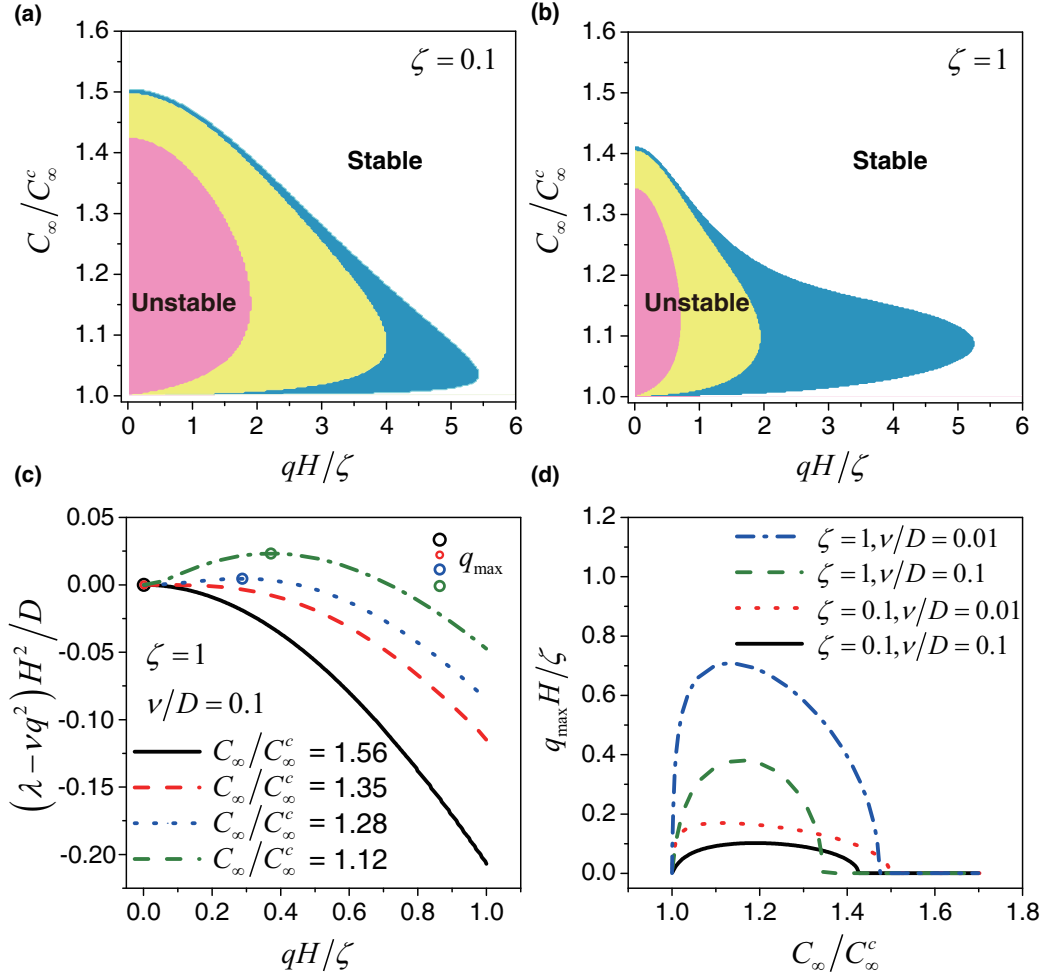}
\caption{Stability of a growing biofilm front as a function of $C_\infty$ and $q$. (a) and (b) The unstable regions correspond to ${\nu /D=0.1}$ (pink), ${\nu /D=0.01}$ (plus yellow), and ${\nu /D=0.001}$ (plus blue). (c) Functional dependence of $\lambda  - \nu   {q^2}$ on $q$. $\lambda  - \nu   {q^2}$ peaks at $q_{\max }$ for given parameters of ${\nu /D}$, $\zeta$ and $ C_\infty$. (d) In the stable regions (compared with (a) and (b)), ${{q_{\max }} \approx 0}$.}
\end{figure}

Since ${\alpha _ +(q=0) } =V/D$, ${\alpha _ -(q=0) }=\lambda _1$ and $\lambda (q=0)= 0$, using perturbation analysis, we find the following asymptotic behavior when $q \approx 0$:
\begin{equation}
\lambda (\mathbf q) - \nu {q^2} \approx \left[\frac{{1 - {\nu  \mathord{\left/
 {\vphantom {\nu  D}} \right.
 \kern-\nulldelimiterspace} D}}}{{{1 \mathord{\left/
 {\vphantom {1 {f({C_\infty })}}} \right.
 \kern-\nulldelimiterspace} {f({C_\infty })}} - 1}} - {\nu  \mathord{\left/
 {\vphantom {\nu  D}} \right.
 \kern-\nulldelimiterspace} D}\right]D{q^2},
\end{equation}
where
\begin{equation}
f({C_\infty }) \equiv {{{\lambda _1}^2\zeta } \mathord{\left/
 {\vphantom {{{\lambda _1}^2\zeta } {{H^2}}}} \right.
 \kern-\nulldelimiterspace} {{H^2}}} - H{\lambda _1}\left(\frac{2}{{1{\rm{ + }}{{{\lambda _1}^2\zeta } \mathord{\left/
 {\vphantom {{{\lambda _1}^2\zeta } {{H^2}}}} \right.
 \kern-\nulldelimiterspace} {{H^2}}}}} - 1\right)\left(\frac{{{C_\infty }}}{{{N_u}\rho }} - 1\right).
\end{equation}
When $C_\infty \ge C_\infty ^c$, $f( {{C_\infty }} )$ is a monotonically decreasing function of $C_\infty $, with $f({C_\infty } = C_\infty ^c) = 1$ and $f({C_\infty } \to \infty ) =  - 1$. So if $\nu/{D} < 1$, for equation $f( {C_\infty } ){\rm{ = }}\nu/{D}$, there is only a single root, and we can denote it as $C_\infty ^P$. Then, one finds that the biofilm surface is stable for $C_\infty \ge C_\infty ^P$ (Region I, Figure 1(b)), while it is unstable for $C_\infty ^c < {C_\infty } < C_\infty ^P$ (Region II, Figure 1(b)). On the other hand, if $\nu/{D}> 1$, the growing front is always stable since $ \lambda (\mathbf q) - \nu {q^2}<0$ for all $q$ (Region IV, Figure 1(b)).

The perturbative calculation near $q=0$ gives us an analytical insight into the condition of instability at the largest length scale across the biofilm. However, Figures 3(a) and (b) show the instability persists up to a finite threshold in $q$, so it will be important to examine the fastest growing mode which corresponds to the maximum growth rate in $q$ space. Using a numerical solution of Equations (20) and (21), we have calculated the overall growth rate as a function of $q$, as shown in Figure 3(c), with the dependence of $q_{\rm max}$ on the nutrient concentration shown in Figure 3(d). It is intriguing to find out that the characteristic length scale of the growing pattern $2 \pi/q_{\rm max}$ exhibits such extreme sensitivity to the nutrient concentration in a narrow range, and disappears when the nutrient concentration is higher than the initial growth threshold by only only 30\%-50\% in our typical examples.

\section{Transient growth behaviour}

When $C_\infty \le C_\infty ^c$, in the approximation that $L \approx \infty$, the growing front eventually stops (equation (9)), yet we can identify the shape of the front by analyzing the transient behavior before it stops. To study the transient behavior, we apply numerical studies on the growth process using difference equations converted from equations (1) and (4). In the simulation (Figure 4), we assume $H/b$ to be a constant and use ${H}$ as the unit length in the $z$-direction, while we define $ \tau  \equiv {H^2}/D $ as the unit time interval. For convenience, define the following dimensionless variables: ${L^{\rm{*}}} \equiv L/H$, ${L_H}^* \equiv L_H/H$, ${t^*} \equiv t/\tau$, ${V^*} \equiv V \cdot \tau/H$, ${c}^* \equiv \frac{{{c}}}{{N_u}{\rho }}$ and ${C_\infty }^* \equiv \frac{{{C_\infty }}}{{N_u}{\rho }}$.

\begin{figure}[htb!]
\centering
\includegraphics[width=1\linewidth]{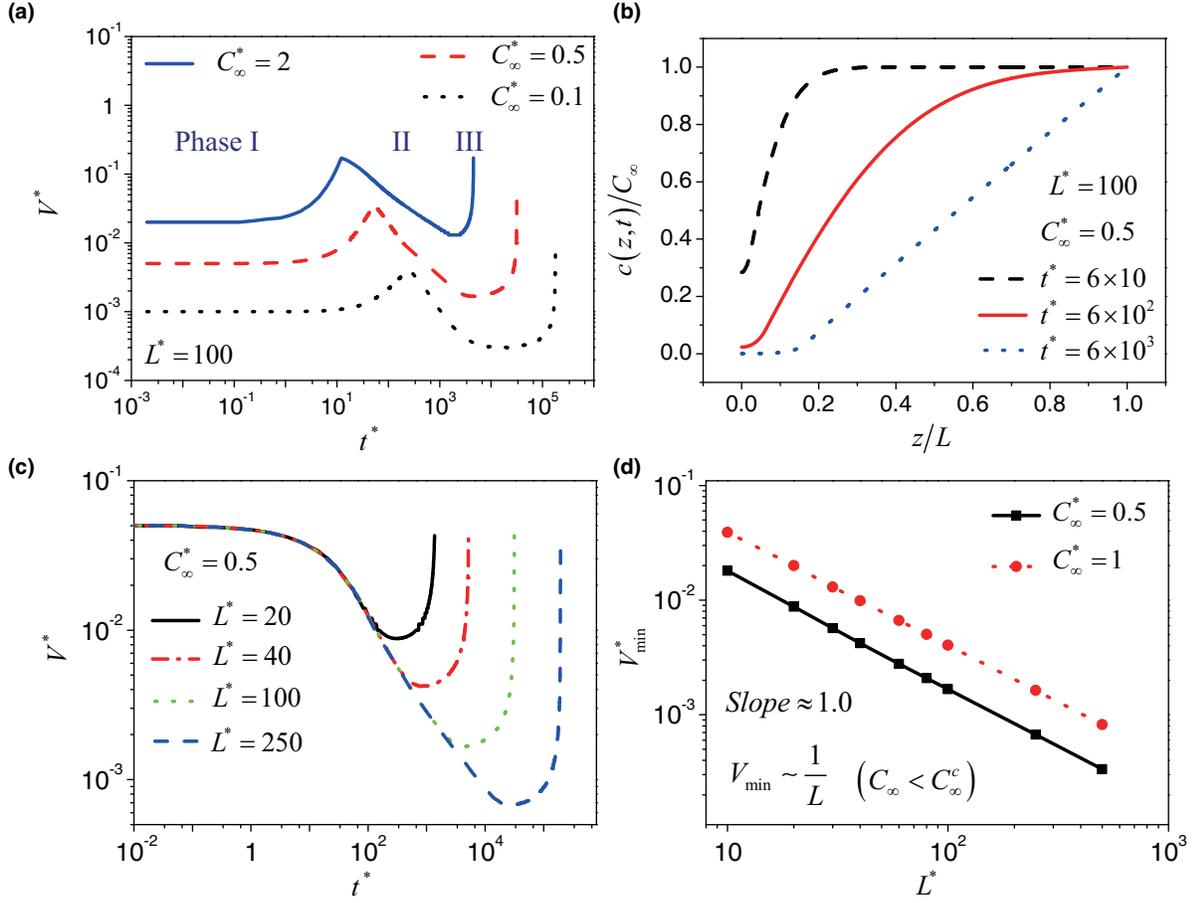}
\caption{Numerical simulation of biofilm growth.(a) Temporal evolution of the velocity for different $C_\infty ^*$ ($C_\infty ^* = 2$ is utilized for growing phase partition). (b) The spatial profile of nutrient at different time points. (c) Temporal evolution of the velocity for different system sizes ${L^*}$. (d) Dependence of ${V_{\min }}$ on ${L^*}$. ($\zeta {\rm{ = 0}}.1 $ and $C{_\infty ^{c*}}\left( {L \to \infty } \right) \approx 3.69$ in (a)-(d); ${{{L_H}^*(t=0)} = 0.1}$ in (a) and ${{{L_H}^*(t=0)} = 1.1}$ in (b)-(d)).}

\end{figure}

We consider the case that a biofilm grows from a thin layer (e.g ${{{L_H}^*(t=0)} = 0.1}$) towards the nutrient source. As is shown in Figure 4(a), the growing front speeds up at first owing to incorporation of more layers of bacterial growth until $L_H \approx H$ (denoted as Phase I). Then the translation speed decreases as the consumption of nutrient by the bacteria overwhelms the supply of diffused nutrient from the source $(z=L)$ (denoted as Phase II). Finally, when $L_H$ approaches $L$, the front speed recovers since nutrient supply is increased in the region near the source $(z=L)$ (denoted as Phase III).

To study the transient behavior before the growing front stops, we ignore the growth process of Phase I by setting ${{{L_H}^*(t=0)} = 1.1}$ (so that $L_H(t=0)>H$). The evolution of the nutrient profile at representative moments is shown in Figure 4(b), where the quasi-steady state nutrient profile (Figure 4(b), blue dot line) agrees with equation (7). To analyze the behavior of $L  \to  \infty$, we simulate the growth process with different system sizes $L$. ${V_{\min }}$, defined as the minimum value in the velocity profile, decreases inversely with system size (Figure 4(c-d)), i.e. ${V_{\min }}\sim1/L$ (Figure 4(d)). In fact, when the velocity is below a threshold, the moving front actually stops (see appendix A for details), and thus there is no Phase III when $L$ is large.

We next focus on the evolution process of Phase II. The stability of the growing front is determined by the local nutrient concentration around the biofilm surface. In the initial state, $c(x,y,z)=C_\infty$ and ${{{L_H}^*(t=0)} = 1.1}$, the local nutrient concentration around the growing front is much higher than that of the steady state (equation (7)). As an approximation, we use the time dependent velocity $V(t)$ in this case as a quasi-steady state quantity to measure the local nutrient adequacy around the growing front. From equation (9) (and Figure 2), one finds that there is a bijective mapping relation between $V$ and $C_\infty$ when $V>0$, and thus we obtain ${V^P} \equiv V\left( {C_\infty ^P} \right)$ and ${V^c} \equiv V\left( {C_\infty ^c} \right)$, the mapping velocity of ${C_\infty ^P}$ and ${C_\infty ^c}$ in a quasi-steady state. As $V(t)$ measures the local nutrient adequacy, so the growing front is stable when $V(t)>V^P$, while unstable when $V^P>V(t)>V^c$. If $\nu/D>1$, the growing front is always stable ($\forall q$, $ \lambda (\mathbf q) - \nu {q^2}<0$) before it stops, thus the biofilm surface is flat (Region V, Figure 1(b)). However, if $\nu/D<1$, ${V^P}= V\left( {C_\infty ^P} \right)>0$, since $V(t\to \infty)=0$ and velocity decreases with time in Phase II, then we can find times $t^P>0$ satisfying $V(t)<V^P$ when $t>t^P$. Meanwhile ${V^c}= V\left( {C_\infty ^c} \right)=0$, thus when $t>t^P$, $V^P>V(t)>V^c$. Consequently, the growing front is unstable before it stops, resulting in a rough surface (Region III, Figure 1(b)).

\section{Discussion}

The growth and patterns formed by a biofilm is summarized in Table 1. According to the mathematical model developed here, in theory, there are five distinct regions. If $\nu /D >1$, the biofilm surface is always flat (Region IV and V), yet the growth is transient when ${C_\infty } < C_\infty ^c$ (Region V), while sustainable when ${C_\infty } > C_\infty ^c$ (Region IV). In reality, the value of $\nu /D$ is usually significantly smaller than one, so these regions (Region IV and V) may be difficult to observe in experiments. If $\nu /D <1$, the sustainable growth threshold is determined by the nutrient threshold $C_\infty ^c$, while the pattern formation is governed by the nutrient threshold $C_\infty ^P$ ($C_\infty ^P>C_\infty ^c$). These two nutrient thresholds, obtained naturally from our analytical analysis, can illustrate the origin of thresholds for roughness and branching in the colony patterns of a recent simulation study \cite{RN334}. Furthermore, the patterning in Regions I-III agree well with those of microbial colonies in the experimental studies \cite{RN333,RN346,RN347,RN355,RN356}.
\begin{table}
\centering
\caption{\label{tab:example}Behavior of the growing front of biofilm.}
\footnotesize
\begin{tabular}{ccc}
\br
Region & Definition  & Behavior\\
\mr
I & ${C_\infty } > C_\infty ^P$, $\nu /D <  1$ & $V>0$, Flat\\
II & $C_\infty ^P > {C_\infty } > C_\infty ^c$, $ \nu /D <  1 $ & $V>0$, Rough\\
III & $C_\infty ^c > {C_\infty }$, $\nu /D <  1 $ & $V=0$, Rough\\
IV & ${C_\infty } > C_\infty ^c$, $\nu /D >  1 $ & $V>0$, Flat\\
V & ${C_\infty } <C_\infty ^c$, $ \nu /D >  1 $  & $V=0$, Flat\\
\br
\end{tabular}\\
\end{table}
\normalsize
To summarize, our study provides a mathematical framework to differentiate growth and patterning behaviors of biofilms and so provides insight for understanding microbial growth.

\ack
This work was supported by the Human Frontier Science Program RGP0061/2013. We thank Berenike Maier and Tom Cronenberg for helpful discussions.

\appendix

\section{Nutrient threshold for growth}
Bacteria are living systems out of equilibrium. When the living environment is harsh, for instance, when nutrients are insufficient, some species of bacteria switch to a protective state named a spore, which is quasi inanimate with significantly lowered energy dissipation (so that the bacteria may survive for even hundreds of years in the harsh environment) \cite{RN362,RN363,RN364}. Definitely, there is a minimum threshold of nutrient flux to initiate cell growth, and we can define it as $\varepsilon$. To consider this effect, the formula of front velocity changes as:
\begin{equation}
V = \frac{1}{{{N_u}}}\int_{{L_H} - H}^{{L_H}} {dz\left[ {{k_0}c(z) - \varepsilon } \right]\theta \left( {{k_0}c(z) - \varepsilon } \right)} ,
\end{equation}
where we have applied $L_H(t=0)>H$. $\varepsilon$ is small and can be neglected when there are enough nutrients to sustain bacterial growth, while it needs to be taken into consideration when the moving speed approaches to zero. For convenience, we can approximate (A.1) as
\begin{equation}
V = \theta \left( {\frac{1}{{{H}}}\int_{{L_H} - H}^{{L_H}} {dz \,{k_0}c(z)}  - \varepsilon} \right)\frac{1}{{{N_u}}}\int_{{L_H} - H}^{{L_H}} {dz\,{k_0}c(z)}  ,
\end{equation}
%where $\varepsilon ' = \frac{H}{{{N_u}}}\varepsilon $.
Using (A.2), we find that there is a threshold of velocity for biofilms growth: when $V <  \frac{H}{{{N_u}}}\varepsilon$, $V =0$. Thus, when ${C_\infty } < C_\infty ^c$, as we find in Figure 4(c-d), ${V_{\min }}$ depends inversely on the system size $L$. When $L$ is sufficiently large, so that ${V_{\min }}< \frac{H}{{{N_u}}}\varepsilon$, then there is no Phase III in the growth process.

Furthermore, we  can derive a solution to the nutrient profile in the steady state ($t\to\infty$) when ${C_\infty } < C_\infty ^c$  (note that $V(t\to\infty)=0$). Supposing that the growing front finally stops at $L_H(t \to \infty)=L-L'$, and using coordinates $(x',y',z',t')=(x,y,z-L+L',t)$, we obtain the following equation as $t' \to \infty$:
\begin{equation}
- D\partial _{z'}^2c'\left( {z'} \right) + \rho {k_0}  c'\left( {z'} \right)\theta \left( { - z'} \right) = 0.
\end{equation}
For $z' > 0$, ${C^ + }(z') ={C_\infty } \left( {A_c} + {B_c} {\lambda _1} z \right)'$; whereas $z' < 0$, ${C^ - }(z') = {C_\infty }{C_1} \,\exp\left(z'\sqrt {\frac{{\rho {k_0}}}{D}} \right)$. The boundary conditions are:
\begin{equation}
\left\{ \begin{array}{l}
c'(z' = L') = {C_\infty }\\
{C^ + }(z' = 0) = {C^ - }(z' = 0)\\
{\partial _{z'}}{C^ + }(z' = 0) = {\partial _{z'}}{C^ - }(z' = 0)
\end{array} \right..
\end{equation}
In steady state $V=0$, which means [see (A.2)]
\begin{equation}
\frac{1}{{H}}\int_{ - H}^0 dz'{{k_0}c'\left( {z'} \right)}  < \varepsilon.
\end{equation}
Combined with (A.4), we obtain ${C_1} =A_c=B_c=1/(1 + {\lambda _1}L')$. When $L$ is large, and $\frac{{L - L'}}{{L'}} \ll 1$, we can approximate $L'$ as $L$ and thus obtain the nutrient profile in equation (7).

When is $L$ large enough to be approximated as $\infty$? The criterion lies in (A.5), which requires a threshold value of the system size for the approximation of $L \to \infty$. Specifically, if we take $L_0$ as the threshold, then
\begin{equation}
\begin{array}{l}
{L_0} = \frac{{D{C_\infty }}}{{{H}\rho \varepsilon}}\left[1 - \exp\left( - H\sqrt {\frac{{\rho {k_0}}}{D}}\right)\right] - \sqrt {\frac{D}{{\rho {k_0}}}} \\
 \;\;\;\;\;\approx \frac{{D{C_\infty }}}{{{H}\rho \varepsilon}}\left[1 - \exp\left( - H\sqrt {\frac{{\rho {k_0}}}{D}}\right)\right],
\end{array}
\end{equation}
 where $L_0$ exhibits an inverse dependence on $\varepsilon$. When $L > L_0$, we can approximate $L$ as $\infty$.

 \section{Upper bound for the growth speed}

We assumed that the uptake rate of nutrient by bacteria takes the Michaelis-Menten form, i.e.
 \begin{equation}
k(c') = k'\frac{c'}{{c' + {K_m}}},
\end{equation}
with the maximum value of $k(c)$ to be $k'$. From equation (3), we find that there is an upper bound to the translation speed, i.e.
\begin{equation}
V \le \frac{{k'H}}{{{N_u}}}.
\end{equation}
Despite the existence of an upper bound to velocity, it is easy to find that the bijective mapping relation between $C_\infty$ and  $V(t \to \infty)$ still holds for $C_\infty \ge C_\infty^c$. Using the same analysis, we can get similar growth and patterning behavior of a biofilm when we apply the correction that arises due to Equation (B.2).

\section*{References}
\bibliography{iopart-num}

\end{document}